\pdfoutput=1
\documentclass[a4paper]{article}
\usepackage[margin=2cm]{geometry}
\usepackage{graphicx}  % allows for indexgeneration

\title{From Procedures, Objects, Actors,
	Components, Services, to Agents
	-- A Comparative Analysis of\\
	the History and Evolution of
	Programming Abstractions}
\author{Jean-Pierre Briot${^\dagger}$}

\date{$^\dagger$ Sorbonne Universit\'e, CNRS, LIP6, F-75005 Paris, France\\
		{\tt\small Jean-Pierre.Briot@lip6.fr}}

\begin{document}

\maketitle
%\maketitlepage

%\begin{abstract}
\abstract{The objective of this chapter\footnote{This preprint has been published
	as a chapter of a book about the French school of programming,
	coordinated by Bertrand Meyer \cite{comparative:analysis:abstractions:fsp:book:2022}.}
is to propose
some retrospective analysis of the evolution of programming abstractions,
from {\em procedures}, {\em objects}, {\em actors}, {\em components}, {\em services}, up to {\em agents},
%have some compare concepts of software component and of agent (and multi-agent system),
%The method chosen is to
by replacing them within a general historical perspective.
% of the evolution of programming abstractions
%(from procedures to objects, actors, components, services, agents\ldots).
Some common referential with three axes/dimensions is chosen:
{\em action selection} at the level of one entity,
{\em coupling flexibility} between entities,
and {\em abstraction level}.
We indeed may observe some continuous quest for higher flexibility
(through notions such as {\em late binding}, or {\em reification} of {\em connections})
and higher level of {\em abstraction}.
Concepts of components, services and agents have some common objectives
(notably, {\em software modularity and reconfigurability}),
with multi-agent systems raising further  concepts of {\em autonomy} and {\em coordination},
notably through the notion of {\em auto-organization} and the use of {\em knowledge}.
We hope that this analysis
%helps at highlight
helps to highlight
some of the basic forces
motivating the progress of programming abstractions
and therefore that it may provide some seeds for the reflection about future programming abstractions.}
%We will conclude at suggesting some prospective directions,
%including some cross-fertilization of current experiences.
%
%between software components and multi-agent systems.
%%
%Meanwhile, we believe that the concepts and technology of software components
%may help at the construction of multi-agent systems.
%We will present in the second part of the article some prospects for cross-fertilization between software components and multi-agent systems.
%\end{abstract}

%\keywords{analysis, comparison, history, programming, software component, agent, multi-agent system, object, actor, service, abstraction, evolution, invocation, binding, coupling.}

\section{Introduction}
\label{section:introduction}

Object-oriented programming, software components
%\cite{componentssei}
and multi-agent systems
%\cite{livresma01}
%,ferberlivresma95}
are some examples of approaches for software design and development with significant impact. 
Both offer abstractions for organizing software as a combination of software elements,
with a common objective of facilitating its {\em evolution} (first of all, replacement and addition of elements).
%We consider that multi-agent systems still push back further the level of abstraction
%and coupling flexibility  between components,
%in particular with the help of their capacities for self-organization and use of knowledge.
%However, we believe that the concepts and technology of software components
%can also help in the construction of multi-agents.
%
%As a first step, and the main part of the article,
In this chapter, our initial objective is to conduct a comparative analysis between software components
and multi-agent systems
(in the following,
%	simplify terminology,
we will use terms, respectively, {\em components} and {\em agents}).
In order to better compare them,
we replace them within some general {\em historical perspective}
of the {\em programming evolution}
(taking some inspiration from \cite{briotinterviewgasserieeeconcurrency98}).
%
%Secondly, we will discuss the potential for cross-fertilization between components and multi-agent systems.
%We will first discuss the contribution potential of agents to components: to design applications based on more autonomous and flexible components,
%for example by using techniques mapping for assembly assistance.
%Then we look at the potential contribution of components to agents as a structure of construction, integration and deployment, not only system-wide multi-agent system,
%but also at the level of an agent.

\section{Related Work}
\label{section:related:work}

There are various comparative studies between agents (and multi-agent systems) and,
e.g., objects \cite{odellobjectsagentsjot02},
concurrent objects \cite{obcp2daidaibook92,briotinterviewgasserieeeconcurrency98}
and actors \cite{introductionseriesactorsagentsieeeconcurrency98}.
This article integrates some of these analyzes and complements them with the concepts of
components and of services, which, to our knowledge, has not yet been the subject of
such systematic comparative studies.
%Note that
Let us mention
%the initiative about the relations between components
%and multi-agent systems
the organization in France in 2004 and 2006
of two successive workshops about multi-agent and components:
%Workshop
``Journ\'ees multi-agents and composants'' (JMAC),
followed by a journal special issue \cite{composantsmaobjet06}.
%Also, note that this chapter is an adaptation and revision of an original article in french
%	published in a french journal
%\cite{composants:agents:tsi:14}.

Let us also cite here, for additional information,
some comparative analyzes about different component models
\cite{classificationcomponentscrnkovicieee11,componentslauieee07}
and about various multi-agent platforms and languages
(based on object-oriented, logic or component-based models)
%\cite{multiagentprogrammingspringer05,
\cite{surveyplatformsinformatica06}.

\section{Analysis}
\label{section:analysis}

%In order to better compare the concept of agent (and of multi-agent system) with the concept of software component, 
%we have chosen to replace them in a general perspective of the evolution of programming, by inspiring us thus in particular of \cite{briotinterviewgasserieeeconcurrency98}.
We have chosen a common conceptual {\em frame of reference} with {\em three dimensions}
that we consider important issues in programming and software:

\begin{itemize}

\item {\em selection of the action} to be performed by an entity --
This is about {\em when} and {\em how} an
%-- or physical,
%	in the case of a robot --
%	for example turn).}
{\em entity}
(a software entity, such as a procedure, a function, an object, an agent, etc.,
or a physical entity, such as a robot or an interconnected device)
will {\em select} (decide) what action to be performed,
through the activation of a corresponding code.
The evolution of programming shows the need
for deferring always {\em later} and {\em further} this decision (this has been coined as ``{\em ever late binding}'').
In addition, for an agent, such a decision may be based, not only on the nature of the invocation,
as for classical programming languages,
but also on the agent's own {\em knowledge} and context
(e.g., by its {\em goals}), in a {\em proactive} and not only {\em reactive} manner;

\item {\em flexibility of the coupling} between entities --
This represents the ability to put in {\em relation} several software entities.
The evolution of programming shows the need to represent
and manipulate such relations independently of the implementation of the entities,
in order to favor {\em adaptability} through some {\em explicit} manipulation of the relations.
%
%% dynamicity -> dynamism ??
%
%offer an architectural independent of the implantation
The concept of {\em software architecture} \cite{softwarearchitecturesbook},
i.e. the {\em assemblage} of components via explicit {\em connectors},
represents therefore a major advance.
The concept of {\em service} brings further {\em dynamism}
(via the concept of {\em discovery} of services)
and {\em autonomy} for the entity itself (the selection of the actual service(s)).
Multi-agent systems raises the description of the coupling
even further
%a step further and higher
%the {\em reification}\footnote{Reification is the process by which an abstract concept about a computer program
%	is turned into an explicit entity created in the programming language.
%	In other words, something that was previously implicit and unexpressed is explicitly formulated and made available to inspection and manipulation,
%	thus often coined as ``making something a first-class citizen''.
%	An example is the Smalltalk programming language which reifies various types of program and implementation entities,
%	such as messages, contexts, classes,
%	as actual Smalltalk objects.
%%	Permanently reified vs transiently reified (in case of errors). Reflexion or reversion.
%	The Lisp programming language has been a true pioneer too, with its uniform vision of considering programs as data.}
%%	The inverse operation, making a reified information into an actual implementation is usually named {\em reflection}, }
%of the architecture and of the discipline of interaction
%a step further and higher,
through concepts such as {\em organization} and {\em interaction protocol}.
%(coordination, breakdown position, negotiation ...)
%using knowledge (organizations, tasks, plans, protocols\ldots);

\item {\em level of abstraction} --
This represents the expression level offered to the designer and to the programmer.
We can observe a progressive quest for higher-level abstractions,
from the initial low-level concepts of {\em instruction},
to abstract concepts of {\em procedure} and {\em abstract data types},
which turn out independent of an implementation platform,
and finally up to {\em knowledge} concepts, such as {\em plan} and {\em intention},
upon which automated reasoning mechanisms can be applied.

\end{itemize}

It should be noted that these three dimensions are not completely independent:
action selection may have some impact on coupling flexibility,
and the choice of abstractions and mechanisms for action selection and for coupling
are clearly related with the level of abstraction.
% in general.
In addition, it is possible
(as,
e.g., in \cite{ghezzioutlookse06})
%	to allow initially symbolic references being later on resolved to become actual memory references.}:)
to consider action selection and coupling
uniformly, both based on a single mechanism:
{\em binding}\footnote{Note that
%dynamic binding (named
	{\em dynamic linking} \cite{dynamic:linking:75}
	was probably introduced as early as in 1959 by the Multics multiuser operating system,
	in order to allow resolving external references at runtime.},
%through a level of indirection.},
which encompasses both:
a) binding of the call to the effective code, in the case of action selection,
and b) binding of a
%link
reference to another entity, in the case of coupling.
However, we prefer to distinguish them,
because their corresponding levels are conceptually distinct ({\em micro} versus {\em macro}),
% vision),
as well as their corresponding professions (programmer versus system architect).
%and their corresponding abstractions and mechanisms
%(e.g., agent architecture versus interaction protocol).

\begin{figure}
\begin{center}
\includegraphics[width=12cm]{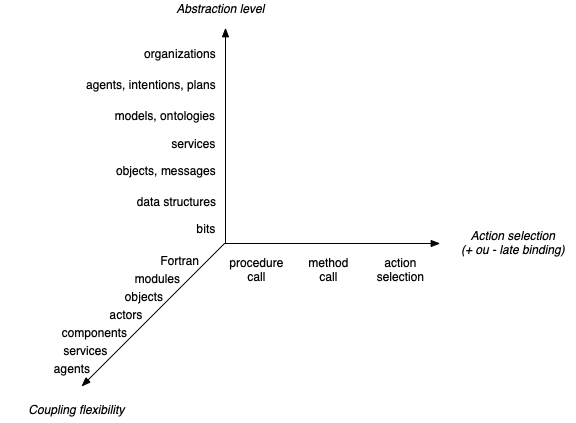}
\caption{Programming evolution}
\label{figure:evolution:3}
\end{center}
\end{figure}

Figure~\ref{figure:evolution:3} illustrates our proposed 3-axes frame of reference.
Each axis will be analyzed, respectively, in
Section~\ref{section:selection:action} (action selection),
Section~\ref{section:flexibilite:couplage} (coupling flexibility)
and Section~\ref{section:abstraction} (abstraction level).

\section{Action Selection}
\label{section:selection:action}

The first programming languages,
%and first of all
e.g., the first version of Fortran,
consider program behavior (code) and program state (data) within a common {\em global data space}.
The different instructions are identified  through their {\em line number}.
The selection of the action (to be performed) is therefore expressed {\em globally} and {\em statically}.

{\em Structured} or {\em modular} programming languages, such as Pascal and then Modula,
introduce some {\em modularization}
%and encapsulation
of the code,
expressed under the form of {\em procedures}.
The selection of the action therefore gains in abstraction,
the indication of the code to be executed being expressed via a {\em symbolic name}
and no longer by a line number.
However, the association of a name of a procedure to its corresponding code remains {\em static}.
%statically determined.
%In parallel,
In some dual movement,
data gradually gains structure and generality,
thanks to the concept of {\em abstract data structures}.
%while remaining however independent
%and external to the procedures.

{\em Object-oriented} programming languages,
with pioneers such as Simula~67 and then Smalltalk,
bring some major innovation,
through the reunion of some procedures and their associated data into a {\em self-contained} capsule,
named an {\em object}.
Data thus become {\em internal} and {\em private} to the object and its procedures (called {\em methods})
and {\em message sending} is the only way to invoke an object,
which will activate one of its procedures.

Some decisive advance is the discipline of {\em late binding} such as in Smalltalk,
i.e. the procedure to be invoked will be determined according to the {\em class}\footnote{A {\em class}
	is the definition of a family of {\em similar} objects.
	It is the class which defines the {\em methods} (procedures) and the {\em variables} (data model)
	common to the objects which will be its {\em instances}, i.e. created by/from it.}
of the actual object invoked,
and not according to the declaration of the {\em type} of the variable that references it\footnote{We deliberately
	do not discuss here the relations between {\em binding} and {\em typing} (and sub-typing),
	due to the fact that they are subtle and
	non consensual.
	For one analysis (among others), see, e.g., \cite{castagna:covariance:acm:95}
	and also the companion chapter by Giuseppe Castagna on types.}.
%	that there are several schools.}.
%.
This means that the binding of the procedure,
and therefore the selection of the action,
is delayed at {\em runtime} and not statically resolved at compile time,
such as in C++ early binding discipline.
(Actually, C++ introduced virtual functions to partially alleviate this limitation,
but this partial solution cannot benefit from further redefinitions of methods once the library has been compiled,
therefore in opposition to Bertrand Meyer's open-closed principle \cite{meyer:oosc:88}).

%%From the point of view of action selection,
%%software {\em components} do not innovate not really because, as we will see in Section~\ref{section:flexibilite:couplage},
%%they tend to focus on the envelope and therefore the coupling rather than on the implementation of the internal wear.
%%They also
%Software {\em components}
%introduce the concept of ``ready to wear, to deploy, and to use''.
%%A component has attributes that can be configured,
%%the analog of the assignment of attributes of an object instance of a class.
%As opposed to an object, which would be orphan and potentially inoperative without its class
%as well as its parent class (superclass) hierarchy
%(e.g.,
%in the case of an object migrating to another site
%%which
%that
%would not have already loaded its associated class hierarchy),
%a component is {\em self-contained}, with all its code
%and also its documentation
%%\cite{composantsvuibert}.
%\cite{szyperski2002component}.
%Therefore,
%%the re thus still has a slight impact on the selection of the action,
%%which,
%on the contrary of an object,
%a component does not require any additional external information
%%external to the component.
%in order to select and process an action.

The concept of {\em agent} introduces {\em internal autonomy} to the selection of the action.
It is no more governed only externally by the nature of the request,
as for a procedure or method call,
but also {\em internally} by the internal state of the agent,
%or more exactly by its {\em knowledge}\footnote{The {\em knowledge} of an agent may be defined as
%	what an agent {\em knows} about its {\em world}
%	(including itself and other agents),
%	this information being described through {\em interpretable} {\em concepts}
%	(i.e., with the potential to be able to {\em reason} about them).},
since this may include be {\em cognitive} information of the agent such as its own {\em goals}.
Therefore, an agent is no longer only {\em reactive} (to invocations) like objects,
but also {\em proactive} \cite{odellobjectsagentsjot02}.
Thus, the concept of action selection takes its full meaning,
as for a robot or a human being,
who can arbitrate his own action(s) at any given time,
depending on both his own objectives
%(internal objectives or resulting from past requests)
and on information collected
%at the time
(messages from other agents or/and {\em perceptions} of the {\em environment}).
{\em Arbitration} can be done at a symbolic level in {\em cognitive agents},
e.g., according to the agent {\em intentions}, in an architecture such as BDI \cite{bdi}.
%as opposed to cognitive agents,
{\em Reactive agents}
%as opposed to cognitive agents,
have much simpler, {\em stimulus}-based action response mechanism,
%fairly straightforward to some type of stimulus.
%In this, they come much
close to message response mechanism in object-oriented programming.
%, in response upon receipt of a message.
Note that there is in fact some continuum between cognitive and reactive agents categories,
with hybrid architectures attempting at reconciling and combining the two approaches
(see, e.g., the InteRRaP hybrid architecture \cite{interrap}).
Last, some {\em sub-symbolic} mechanisms
(without an explicit representation of the world) for regulation,
often inspired by biology (metabolism, emotions, motivation, adaptation, see, e.g., \cite{animats})
can also be incorporated to agents.

%Reflecting on the evolution of action selection,
Les Gasser proposed in 1998
as one of the fundamental concepts of agent programming
the concept of {\em structured persistent action},
%to capture the way
in which an agent is {\em autonomously} and {\em persistently} trying to accomplish something,
independently of the way it is programmed
\cite{briotinterviewgasserieeeconcurrency98}.
In standard procedural programming,
the programmer explicitly controls the attempts,
while the concept of structured persistent action
%This notion
abstracts and encapsulates such a mechanism.
More precisely,
the designer provides the description of the objective or criteria for success,
as well as in general a collection of methods and recipes, which the agent will select and control autonomously.
Note that some similar mechanisms have already been proposed,
for instance {\em declarative} programming and {\em backtrack} in logic programming languages
such as Prolog,
%by declarative programming and backtrack,
or the general concept of {\em search}.
But, in our opinion, the concept of structured persistent action
represents in an interesting way the encapsulation of:
a notion of
{\em choice},
{\em information}\footnote{Some information
	about the possible choices of actions related to the domain in which the agent acts.
	Such information can be symbolic (beliefs, models, plans\ldots) or not,
	depending on the choice of agent architecture and of the representation of the world.},
an {\em iterative control structure} (of type repeat until),
and proper {\em resources} (own process or thread).
In addition, we consider the interaction of the agent with its environment
to ensure some {\em feedback} over its actions and choices (e.g., through some reinforcement learning mechanism).
Last, we may observe that the selection (and therefore the choice)
of the action takes place at the moment of the action by the agent
and not at the moment of the programming of the agent.
Therefore, the concept of agent is situated within the quest for ``{\em ever late binding}''.

Table~\ref{table:selection:action}, inspired by \cite{odellobjectsagentsjot02},
summarizes our analysis.
The horizontal axis of Figure~\ref{figure:evolution:3} illustrates the evolution of action selection
within our frame of reference.

\begin{table}
\begin{center}
\begin{footnotesize}
\begin{tabular}{|c||c|c|c|c|}
\hline
{\sl Programming}	&{\it Monolithic}		&{\it Modular}		&{\it Object-oriented}		&{\it Agent-Oriented}\\
				&{\it ex: Fortran}	&{\it ex: Pascal}		&{\it ex: Java}			&{\it ex: AgentSpeak}\\
\hline\hline
{\sl Behavior}		& Global			&Modular			&Modular				&Modular\\
\hline
{\sl State}			& Global			&Modular			&Modular				&Modular\\
				&				&and external		&and internal			&and internal\\
\hline
{\sl Invocation}		&Global			&External			&External				&Internal 	and external\\
{\sl (and Selection)}	&and static		&and static		&and dynamic			&and dynamic\\
				&{\it (goto)}		& {\it (procedure}	& {\it (method}			& {\it (ex: goal-}\\
				& 				& {\it call)}			& {\it call)}				& {\it driven)}\\
\hline
\end{tabular}
\end{footnotesize}
\caption{Structure of entities and action selection}
\label{table:selection:action}
\end{center}
\end{table}

%\begin{figure}
%\includegraphics[width=10cm]{evolution-selection-en.png}
%\caption{Evolution of action selection}
%\label{figure:evolution:selection:action}
%\end{figure}

\section{Coupling Flexibility}
\label{section:flexibilite:couplage}

The modeling of the {\em coupling} between software entities is a fundamental aspect for the structuring of the software.
It actually covers several facets:

\begin{itemize}

\item {\em structure}: the architectural concepts (e.g., {\em references}, {\em connectors}\ldots)
for the structural coupling between software entities;

\item {\em communication}: the modes of communication between software entities,
characterized mainly by: the mode for the {\em designation} of the receiver, the mode for {\em data transfer},
and the mode for {\em temporal coupling}.

\end{itemize}

\subsection{Structural Coupling}
\label{section:couplage:structurel}

The question of the {\em structural coupling} between software entities has been initially addressed
by the notion of {\em reference} to an entity, through some means for identifying it (identifier).
Therefore, one may designate a software entity (e.g., some data, object or function),
%\footnote{{\em Simple data} in early programming languages,
%	{\em functions} in functional programming languages,
%	{\em objects} in object-oriented programming languages\ldots},
in order to use it
%and transmit
and to communicate its reference to other entities.
This model, simple but effective and general, survived with object-oriented programming languages.

For instance, an object {\tt A} references an object {\tt B},
and thus will be able to send requests to {\tt B}.
In practice, the internal representation (implementation) of {\tt A}
includes a variable whose value is the identifier of object {\tt B}.
Changing a reference is easy, by just changing the value of the variable,
for instance to the identifier of a third object {\tt C}.
However, we can observe that this modification can be done only {\em internally} to object {\tt A},
the only one authorized to access its private data (following the encapsulation principle).

\begin{figure}
\begin{center}
\includegraphics[width=9cm]{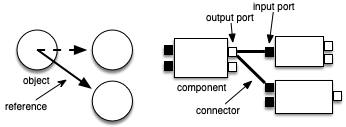}
\caption{Objects coupling versus components coupling}
\label{figure:couplage:objet:vs:composant}
\end{center}
\end{figure}

A serious limitation occurs when we want to {\em extend} a reference,
for instance so that {\tt A} refers {\em both} to {\tt B} and to {\tt C}
(see the left part of Figure~\ref{figure:couplage:objet:vs:composant}).
Since a variable has only one value, this cannot be expressed directly.
It is therefore necessary to introduce some data structure (a collection, e.g., a list),
containing {\tt B} and {\tt C}.
The message sending instruction must also be modified,
by introducing an iterator on the collection.
Overall, this implies the modification of the internal representation of object {\tt A} (in other words, to reimplement it),
whereas it is only a question of extending the reference and the coupling,
initially from {\tt A} to {\tt B},
into from {\tt A} to {\tt B} {\em and} {\tt C}.

The concept of {\em software component}
%\footnote{For a more complete analysis
%	of the characteristics of software components
%	and a comparison between different component models,
%	please
%	see, e.g., \cite{classificationcomponentscrnkovicieee11}
%	or/and
%	and
%	\cite{componentslauieee07}.}
brings some notable improvement to this problem by {\em externalizing} the references,
describing them as explicit {\em output interfaces}.
Therefore,
a component regains some {\em symmetry} at the level of interfaces
between {\em input interfaces} (which are traditional for procedures and objects)
and {\em output interfaces},
alternatively named, respectively,
{\em provided interfaces}
and {\em required interfaces}.
(For a more complete analysis
of the characteristics of software components
and a comparison between different component models,
%	please
see, e.g., \cite{classificationcomponentscrnkovicieee11}
%	or/and
and
\cite{componentslauieee07}.)
%as opposed to, or rather
%in addition to\footnote{In that sense,
%	we can say that a component regains {\em symmetry}, at the level of {\em interfaces},
%	with both input and output interfaces.}
%the traditional {\em input interfaces} for objects and procedures.
%Some alternative terminology often used is {\em required} interfaces (output)
%and {\em provided} interfaces (input) \cite{composantsvuibert,componentslauieee07}.
%The identifier of an interface is usually named a {\em port}.

%The management of the
Coupling thus becomes {\em explicit},
{\em reified} (i.e. coupling is made into first class entities, the {\em connectors})
and {\em external} (to the software entities).
%into {\em connectors}, which will connect ports.
%The
%desired reconfiguration
Previous example
is therefore achieved by the simple addition of a connector,
as illustrated in the right part of Figure~\ref{figure:couplage:objet:vs:composant}.

Note that a component can have multiple interfaces (input or/and output interfaces).
To be able to identify them individually, an identifier, usually named a {\em port}, is associated to each interface.
%and is usually named a {\em port}.
%Each interface is usually identified as a {\em port}.
This is an important difference with an object which has only one identifier and entry point.
An interesting consequence is that components are {\em compositional}.
That is to say that a composition of several components is equivalent\footnote{Actually,
	we should distinguish
	between {\em functional composition},
	which is a simple assembly of components,
	and {\em structural composition},
	which encapsulates a functional composition and identifies it as a new component,
	%\cite{nierstrasz},
	often referred to as a {\em composite component}.
%	See also another terminology, based on binding
%	-- {\em horizontal binding} for the functional composition
%	and {\em vertical binding} for the structural composition --,
%	in \cite{classificationcomponentscrnkovicieee11}.
%	(Another terminology, based on binding, is
%	{\em horizontal binding} (functional composition)
%	and {\em vertical binding} (structural composition)
%	see \cite{classificationcomponentscrnkovicieee11}.)
	\cite{classificationcomponentscrnkovicieee11}
	analyzes their respective binding techniques, named {\em horizontal binding}
%	for the functional composition
	and {\em vertical binding}.
%	for the structural composition.
	We believe that the concept of structural composition is important,
%	\cite{OCM2000},
	as it provides {\em encapsulation} and {\em hierarchy},
	which both proved to be useful to control complexity.
	However, only a minority of component models support composite components
	(e.g., Fractal \cite{fractal} and MALEVA
%	\cite{malevaobjet06}
	\cite{maleva:promas:book:07},
	but not JavaBeans \cite{beans} nor CORBA Component Model (CCM) \cite{ccm}).}
to a component with the corresponding union of input ports and output ports.
On the opposite, objects are not directly compositional:
a composition of several objects is not immediately equivalent to an object,
as it has more than one entry point.

Therefore, components provide an explicit architectural vision.
The notion of {\em software architecture} \cite{softwarearchitecturesbook}
%of an application
focuses on the logic of the coupling between the components,
independently of their internal implementation.
{\em Architecture description languages} (ADL) \cite{softwarearchitecturesbook}
are dedicated to the specification of the architecture of an application
and they are indeed very different from standard programming languages.
Information about the typing of component interfaces is used to verify correctness of the assembly,
i.e. the conformity between the interfaces which are brought in relation.
Different types of connectors are usually considered
and correspond to different {\em architectural styles} (e.g, layered, pipes and filters,
broadcast of events, etc.
\cite{softwarearchitecturesbook})
and their associated communication protocols.
Connectors can also represent {\em non-functional properties}
(such as distribution, quality of service, etc.)
and therefore have their own semantics
\cite{garlanformalconnectors}.

In order to express not only specifications about the types of data (typing information)
but also about the {\em behavior} of components,
notions of {\em contracts} have been proposed.
For instance, \cite{contratscomposants} considers four successive levels of contracts:
syntactic, behavioral, synchronization and quality of service.
Depending on the case, they can be {\em guaranteed}, {\em verified} or {\em negotiated}.
The syntactic level is based on a type system.
The behavioral level is usually based on {\em assertions}
(the three main types being: pre-conditions, postconditions, and invariants).
%The general idea of contracts,
But, compared to the use of assertions within a program,
the idea of contracts is to specify them in a modular way
and {\em visible} through the interfaces of a component,
%outside of the component,
%within the interfaces
in order to be able to specify properties that can engage more than one software entity \cite{programmationparcontratsmeyer}.

In addition, components
introduced the idea of ``ready to wear, to deploy, and to use'',
%A component has attributes that can be configured,
%the analog of the assignment of attributes of an object instance of a class.
i.e. a component is some self-contained {\em unit of deployment}, with all its code
and also its documentation
%%\cite{composantsvuibert}.
\cite{szyperski2002component}.

\begin{figure}
\begin{center}
\includegraphics[width=12cm]{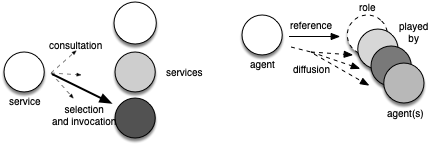}
\caption{Services coupling and agents coupling}
\label{figure:couplage:service:agent}
\end{center}
\end{figure}

The concept of {\em service} of {\em service-oriented architectures} (SOA)
-- including in particular {\em web services} \cite{webservices} --
brings dynamism to coupling, and moreover autonomy,
via {\em discovery} 
and {\em dynamic selection} of other services (as shown in the left part of Figure~\ref{figure:couplage:service:agent}).
Coupling between entities is therefore no longer only managed by the designer of the application,
but by the entities themselves
(this corresponds to some degree of {\em self-organization}).
For instance, an electronic travel agency service,
looking for services to perform subtasks (e.g., flight reservation, hotels, etc.),
will thus be able to {\em identify}, {\em select}
(in general,
according to various criteria, e.g., availability, price, flexibility, etc.),
and {\em contract} sub-services.
Therefore, services are subject to more or less elaborate descriptions, which are made available ({\em published}),
e.g., through some directory of services, similar to telephone numbers yellow pages.
For web services, UDDI (Universal Description, Discovery and Integration)
and WSDL (Web Services Description Language) standards \cite{webservices}
specify, respectively, directories and descriptions of services.
%Coupling between entities is therefore no longer only managed by the designer of the application,
%but by the entities themselves.

Multi-agent systems further extend dynamism and autonomy
%capacity of {\em self-organization} of the coupling
%between entities,
by trading some {\em syntactic} coupling
(following some typing discipline)
for some {\em semantic} coupling,
based on knowledge (via abstractions such as: {\em task}, {\em plan} and {\em intention})
and some {\em social organization of work}
(via abstractions such as: {\em organization}, {\em role}, {\em norm} and {\em negotiation}).
%{\em coordination}
%%{\em decomposition},
%and {\em negotiation}.

%Some fundamental concept,
%more knowledge-oriented level than the concept of software architecture,
%is the
An {\em organization}
%It
specifies the different {\em roles} constituting it
%(for instance,
(e.g., roles of producer, consumer and broker)
and their {\em relationships} (e.g., dependency and hierarchy).
%Some given
A {\em role} can be {\em played} by one or more agents
and the same agent can also possibly play more than one role simultaneously.
%The {\em role} abstraction allows a more generic description of the architecture,
%as well as the relationships and interactions between agents.
%Thus,
%Therefore, some
Note that an agent referencing a role
subsumes a reference to all the agents {\em fulfilling} (at the time of the interaction) this role,
as illustrated in the right part of Figure~\ref{figure:couplage:service:agent}.
(This mechanism
of {\em abstract role designation} of the receiver will be further analyzed in Section~\ref{section:designation:receveur}.)
%In this sense,
%structural coupling is external as for components,
%but gains some degree of {\em implicitness}, unlike explicit connections between components\footnote{However,
%	from an architectural point of view,
%	agents generally have only one entry point, as for objects,
%	which moreover often serve as a support for their location.
%	As a result, they are generally not compositional, unlike components.}.
	
%	with some explicit {\em knowledge},
%through various concepts such as: {\em organization}, {\em task}, {\em plan}, {\em norm},
%{\em coordination}
%%{\em decomposition},
%and {\em negotiation}.

%as well as the means for {\em collaboration} between entities
%(e.g., coordination, decomposition, negotiation\ldots),
%using {\em knowledge} (organizations, tasks, plans\ldots).

Two important capacities of an organization are its {\em dynamism} and its {\em autonomy}.
%(self-organization and self-reorganization).
%An organization is inherently {\em autonomous}  {\em dynamism} and its {\em autonomy}
%({\em self-organization}).
%as it can evolve according to the needs.
Some dynamic reorganization can be triggered:
in a {\em top-down} manner,
e.g., the reorganization of a robotic football team
(as in the RoboCup contest \cite{robocup}),
according to a more defensive strategy
on the initiative of the coach \cite{moise+};
or in a {\em bottom-up} manner,
%at the initiative of the agents themselves,
e.g.,
with the dynamic formation (and then dissolution)
of a micro-organization of type ``one-two''
on the initiative of some player agent \cite{cassiopee}.
Examples of abstract models of organizations are AGR \cite{agr} and MOISE+ \cite{moise+}.

%In addition, as
As for services, multi-agent systems also often use various mechanisms for putting agents into relation:
%dynamic and indirection,
by some intermediary agents,
directory agents,
or facilitator agents
guided by the content of the message (e.g., in KQML \cite{kqml});
%Setting relationship, in particular for the
or by some selecting and contracting mechanism,
%call for proposal {\em selecting} and {\em contracting} agents,
%in view to process tasks,
as, e.g., the {\em contract net protocol} \cite{cnp}
(which will be introduced in Section~\ref{section:synchronisation}).

%{\em Relations} between the agents are therefore very {\em dynamic}
%and in part or completely managed by themselves or through organizations.

To conclude, note that the software architectures and components communities started to
%address
support
%such challenges of
%{\em dynamism and
%capacity for
automatic reconfiguration,
%capacities atomatic of architecture,
e.g., for nomadic applications \cite{dubusautoadaptabilitearchitecturesjc06}.
%However,
But
the knowledge and social-oriented approach
%characteristic of multi-agent systems, remains
%(for the moment at least, see
%Section~\ref{section:assistance:assemblage} below)
%Publi Jeff Kramer et al. ??
%the prerogative of multi-agent systems.
%The multi-agent approach
of multi-agent systems is more ambitious, and therefore also more difficult to {\em verify}.
We thus find out some classic dilemma between
the growing needs for {\em flexibility}, through some delegation of initiative,
and the needs to ensure some {\em guarantees} on the operability of the system.

\subsection{Communication Coupling}
\label{section:couplage:communication}

The expression of the {\em mode of communication} between software entities
includes several important characteristics (sub-facets).
We consider here the three main ones:

\begin{itemize}

\item how to {\em designate the receiver(s)},
e.g., point to point, multi-point, indexed by content, via the environment, etc.;

\item the mode for {\em data transfer},
e.g., unidirectional, bidirectional with value return, via a shared space, etc.;

\item the {\em temporal coupling}
(in other words, the way communications are {\em synchronized}),
e.g., synchronous, asynchronous, with an anticipated (future) response,
coordinated by a protocol, etc.

\end{itemize}

\subsubsection{Designation of the Receiver}
\label{section:designation:receveur}

The mode of communication between objects is fundamentally {\em point to point},
i.e. one to one and with explicit designation of the receiver of the message.
Components introduce {\em multi-point} communication,
as an output of a component can be connected to more than one component.
An interesting type of connector is the event broadcasting connector,
corresponding to the {\em publish-subscribe} architectural style \cite{softwarearchitecturesbook}.
It offers an {\em indirect} and {\em dynamic} management of connections by the components themselves,
through a mechanism of subscription of a component to the event broadcaster.
(The subscription criteria
and the distribution method may vary,
see, e.g., the classification proposed in \cite{eugsterpublishsubscribeacmcs03}.)
This type of mechanism
became widespread
%even outside the world of components
(e.g., in applications based on standard objects)
%but according to us it is
although it remains very representative
of the concept of connector between software components,
defined and manipulated externally to them
(as it has been analyzed in Section~\ref{section:couplage:structurel}).

The {\em shared spaces} (repositories) architectural style,
illustrated by, e.g., blackboards and tuple-spaces
(for instance, the LINDA model \cite{linda}),
introduces a mode of designation of the receiver completely {\em implicit},
since it will be {\em indexed} by the actual {\em content} of the message.
In this model, active entities (e.g., processes or agents)
can insert and index structured data within the shared space.
Data will be consumed {\em opportunistically} by active entities
%waiting for
looking for
the corresponding data patterns.

Services, and moreover multi-agent systems,
generalize mechanisms of indirect and dynamic designation,
through some contracting protocols or the consultation of  broker or directories agents
(as it has been presented in Section~\ref{section:couplage:structurel}).
Services or agents can therefore dynamically select their own {\em interlocutor}.
Some more implicit mechanism is the notion of {\em facilitator},
guided by the  content of the message \cite{introductionseriesactorsagentsieeeconcurrency98}
(e.g., in KQML \cite{kqml},
to be analyzed in Section~\ref{section:abstraction:interoperability:languages}).
Another type is the
%indexing of communications
abstract designation of a receiver
through a {\em role},
as, e.g., in the AGR (agent group role) {\em organizational model} \cite{agr}.
In such role-based models,
agents usually designate some role
(e.g., midfielder or striker, in a RoboCup football organization),
rather than some specific agent, as the receiver of a communication.
As a consequence,
all the agents fulfilling this role at the time of communication will receive the information
(see the right part of
%Figure~\ref{figure:couplage:objet:vs:composant}).
Figure~\ref{figure:couplage:service:agent}).

Last, in certain types of multi-agent systems, in which the environment (physical or not)
is explicitly modeled,
the agents can communicate via the environment,
though inserting specific data, for example {\em pheromones} for ant-based algorithms.
(Such algorithms can be used as a general meta-heuristic optimization method,
%(see, e.g., \cite{agentssitueslivretechnologiessma09}),
the environment having then no longer relation with a physical reality.)
Note that, there is also some trend in multi-agent systems for promoting the environment
as a first-class abstraction
(for more details, see, e.g., \cite{environmentsmas}).
There is also a similar trend
for promoting entities without internal goals and characterized by a function
as first-class entities named {\em artefacts},
which are manipulated (use, selection or construction) by agents \cite{artefacts:jaamas:2008}.

\subsubsection{Data Transfer}
\label{section:transfert:donnees}

The mode for {\em data transfer} in object-oriented programming is {\em bidirectional},
with some {\em return of value}
(unless
the programmer explicitly specifies that there is no return value,
e.g., in Java using the special data type {\tt void} which represents the absence of data).
It is inherited from the procedural or functional call.
It corresponds (as we will see in Section~\ref{section:synchronisation}) to a {\em synchronous} call,
i.e. with the sender suspending its activity
while waiting for the completion of the processing of the request by the receiver.
% (recipient).

The {\em actor} model \cite{HEWITT1977323,aghalivreacteurs} introduces some {\em unidirectional}
(and {\em asynchronous}, see Section~\ref{section:synchronisation}) data transfer mode.
This was motivated
by the concurrent and moreover distributed nature of the model,
in order to avoid unnecessary and unbounded waiting for an acknowledgement
of data transfer completion.
Therefore, data transfer is carried out only {\em one-way} from the sender to the receiver.
If the receiver wants to return a value,
it must be done explicitly, by sending another message.
Some languages based on actors,
as for instance ABCL (Actor-Based Concurrent Language) \cite{abcloopsla86},
provide the programmer with a choice between a
one-way asynchronous message send and a two-way synchronous call.
(Note that the Actalk object-oriented framework
offers in a single pedagogical framework various types of actor-based
and object-oriented concurrent programming abstractions,
regarding various models of:
action selection, activity, communication \cite{actalk:lmo:94}
and internal synchronization \cite{actalk:synchronization:isotas:96}.)
%	\cite{actalk:ecoop:89,actalk:synchronization:isotas:96}.}.

\begin{figure}
\begin{center}
\includegraphics[width=11cm]{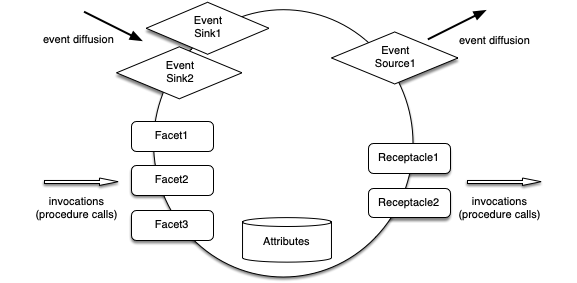}
\caption{CCM component model}
\label{figure:ccm}
\end{center}
\end{figure}

Component models,
such as CORBA component model (CCM) \cite{ccm}\footnote{Note that
	an example of a more recent component model (also an industry standard),
	also integrated into a service-oriented architecture, is CSA (Composite Services Architecture) \cite{websca}.
	However, we have chosen here to illustrate our analysis through the CCM model,
	for its historical and pedagogical value.}
often also propose these two modes of data transfer:
bidirectional though a procedure call (via input and output interfaces,
named {\em facets} and {\em receptacles} in CCM),
and unidirectional though event diffusion (via event {\em sources} and {\em sinks}),
see Figure~\ref{figure:ccm}.

The shared spaces architectural style
%(see Section~\ref{section:designation:receveur})
(presented in Section~\ref{section:designation:receveur})
introduces a mode of data transfer, indirect,
via some mediation structure and the distinction between production and consumption.

Services are generally based on simple invocation protocols,
in particular the case of web services.
%e.g., with the SOAP protocol \cite{webservices}.
One of the main reasons for the success of web services
is likely their easy deployment on top of the widespread web infrastructure and its HTTP protocol.
The SOAP protocol \cite{webservices}
(originally the acronym for ``Simple Object Access Protocol'')
supports both bidirectional and  unidirectional modes.

Multi-agent systems generally offer the unidirectional (and asynchronous) transfer mode of actors
but expressed within more elaborate {\em agent communication languages},
%\footnote{See
%	the paragraph about agent communication languages in Section~\ref{section:abstraction}.}
%	Section~\ref{section:abstraction:interoperability:languages}.},
which allow to specify with precision and details the nature of the information to be communicated
(as it will be presented in Section~\ref{section:abstraction:interoperability:languages}).

Last, some possible communication via an environment
(by adding, removing, or consuming data,
as in the shared spaces architectural style,
such, e.g., the LINDA model \cite{linda},
presented in Section~\ref{section:designation:receveur})
represents some indirect mode of data transfer.

%24 juillet 2022 soir

\subsubsection{Temporal Coupling (Synchronization)}
\label{section:synchronisation}

The original communication mo\-del between software entities
%(originally
(in a sequential and centralized world)
is the procedural or functional call, with return of a value.
The sender activity is {\em suspended} during the processing of the request by the receiver.
A direct transposition into a concurrent setting sticks to these principles,
with the sender waiting for the call to be completed -- this is referred to as {\em synchronous} transmission.
% (message sending).
A direct transposition into a distributed setting
is represented by the RPC ({\em Remote Procedure Call}), also synchronous.

The actor model
%\cite{aghalivreacteurs}
introduces an {\em asynchronous} mode of communication as its foundation,
i.e. without waiting for the message to be processed
-- and before that, to be received -- by the receiver.
Asynchronous communication is more appropriate to a concurrent or/and distributed setting
%(actors being simultaneously active,
%this avoids waiting for the conclusion of treatment by the receiver)
%distributed
(due to the potential latency of the communication network,
this avoids waiting for the delivery of the message to the receiver,
as well as its availability to process it).
Therefore, the actor model assumes the existence of a {\em mailbox} for each actor,
which will store the messages in the order of the arrival (FIFO type discipline).
The actor model thus introduces some {\em temporal decoupling}
between {\em sending}, {\em receiving}, {\em processing start}, and {\em processing completion} of the message.
As explained in Section~\ref{section:transfert:donnees},
some actor-based languages, such as ABCL,
can provide both one-way asynchronous and two-way synchronous communication 
\cite{abcloopsla86}.
Another mode provided by actor languages and ABCL is a promise/anticipation of the response, often named {\em future}.
It corresponds to {\em eager evaluation}
(also coined as {\em wait by necessity} in \cite{caromel:concurrent:eiffel:cacm:93}),
the actual exact opposite of {\em lazy evaluation}.
Scala is an example of a programming language which integrates
functional, object-oriented, and actor programming \cite{bookscala10}.
Last, for an analysis about the different ways of mapping the object-oriented programming model
to concurrent and distributed programming requirements,
please refer, e.g., to \cite{cdoopacmsurveys98} or \cite{opr:tsi:96}.

%Lundi 25 juillet nuit

%Communication between agents inherits the {\em asynchronous} (and {\em unidirectional}) mode of actors.
{\em Agent communication languages} (ACL), in particular FIPA\footnote{FIPA is the acronym for
	the Foundation for Intelligent Physical Agents,
	an IEEE Computer Society standards organization
	which promotes agent-based technology
	and the interoperability of its standards with other technologies \cite{fipa}.}
ACL \cite{fipaacl},
%often 
allow the specification of a protocol associated with a communication.
The {\em protocol} specifies the {\em coordination} of valid message exchanges between agents.
Temporal coupling is therefore expressed in a relatively general manner
and with an arbitrary number of messages and agents.
Example of families of agent protocols are:
interaction (e.g., inform, request, deny\ldots),
coordination (e.g., simple or iterated call for proposals, see next paragraph)
and
auction (e.g., English or Dutch, with, respectively, increasing or decreasing initial price).
%negotiation,

\begin{figure}
\begin{center}
\includegraphics[width=7cm]{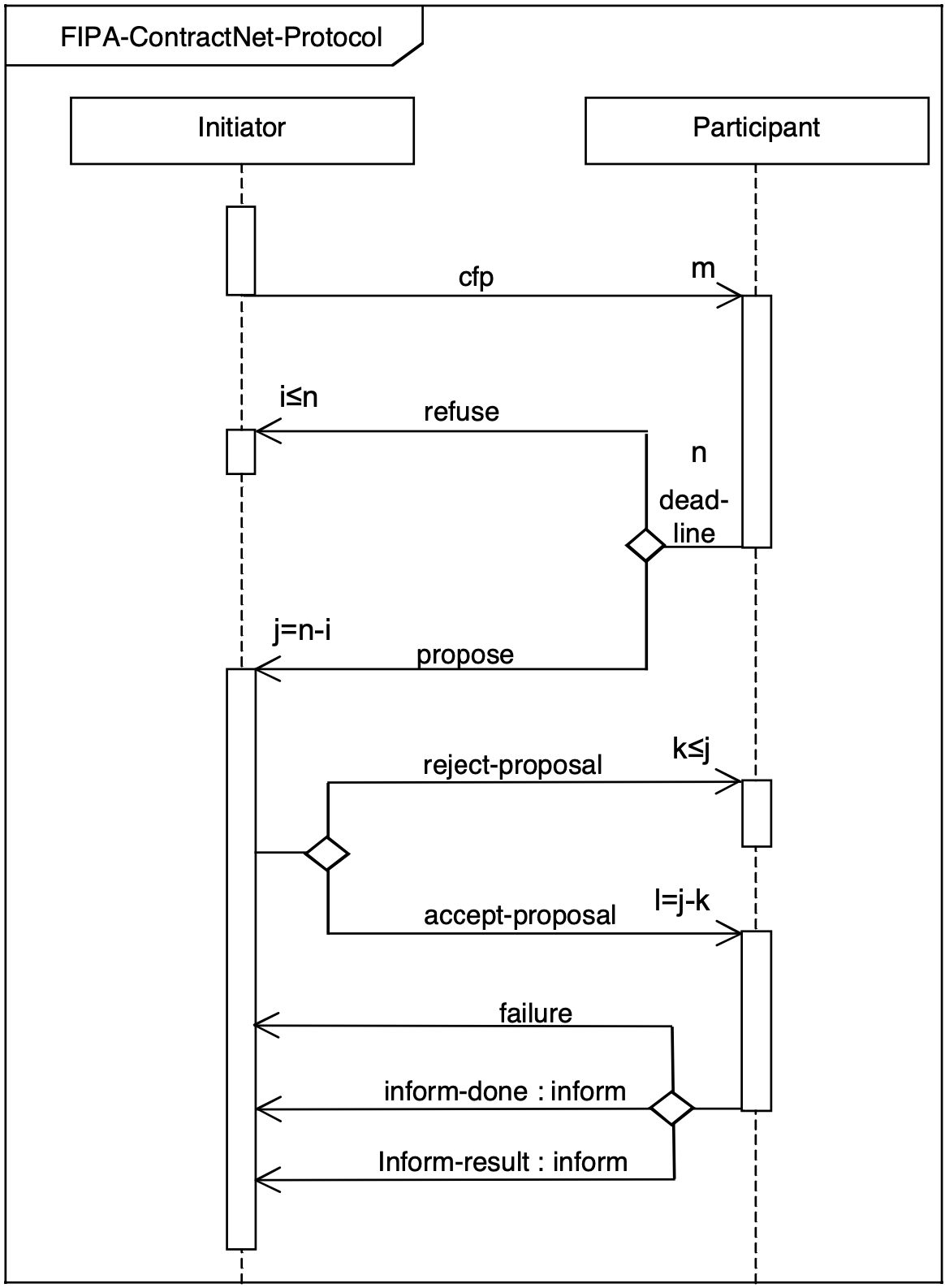}
\caption{Contract net protocol. Figure reproduced from FIPA Contract Net Interaction Protocol Specification, Foundation for Intelligent Physical Agents, 2002.}
%``http://www.fipa.org/specs/fipa00029/SC00029H.html#_Toc26669257''}
\label{figure:cnp}
\end{center}
\end{figure}

A classic example of a multi-agent protocol is the call for proposals
(also named the {\em contract net protocol}).
Figure~\ref{figure:cnp} shows the corresponding interaction diagram (as specified by FIPA \cite{fipaacl}).
Successive phases are:

\begin{itemize}

\item the broadcast of the initial call
(where {\tt cfp} stands for call for proposals)
by the {\em initiator} (also named {\em contractor})
to the participants;

\item various proposals (or refusals) made by the participants,
controlled by some deadline (timeout) for responding;

\item the selection and acceptance (or rejection) of a proposal by the initiator;
and finally

\item
the communication by the selected participant (also named {\em sub-contractor})
about the finalization and the result (or the failure) to process its proposal.

\end{itemize}

Web services also offer analog coordination mechanisms, also named {\em choreography}.
The Web Services Choreography Description Language (WS-CDL) has been initially defined
with this intent by the W3C
(World Wide Web Consortium standard \cite{w3c}).
It has since been replaced by the BPEL (Business Process Execution Language)
and BPMN (Business Process Model and Notation) standards \cite{bpel13}.
(We will not detail here
the characteristics of services and Web services,
which are the subject of standards and numerous technical specifications,
because that would be the subject of another article.
See, e.g. \cite{webservices} and \cite{bookpapazoglouwebservices},
as well as \cite{paynewebservicesagents08} for an agent perspective on web services.)
%	,
%	this one focusing on
%	%in priority
%	the relationships with components and agents.}.

Table~\ref{table:flexibilite:couplage} summarizes the evolution of coupling
according to the 2 main facets: structure and communication,
%and the
%3 sub-facets of the communication facet:
the latter one with its
3 sub-facets:
designation of the receiver(s), data transfer mode, and temporal coupling (synchronization).

The diagonal axis of Figure~\ref{figure:evolution:3} illustrates the evolution of coupling flexibility
within our frame of reference.
Note
that the evolution of coupling flexibility is
not completely linear:
%	nor single directed:
actors have been proposed before components but their respective main focuses
are different (respectively, concurrency and architecture);
web services have been proposed after multi-agent systems.
%	and have benefited from them.}.

\begin{table}
\begin{center}
\begin{footnotesize}
\begin{tabular}{|c||c|c|c|c|c|}
\hline
{\sl Coupling}		&{\it Objects}	&{\it Actors}		&{\it Components}	&{\it Services}	&{\it Agents}\\
				&			&				&				&			&\\
\hline\hline
{\sl Structure}		&Implicit		&Implicit			&Explicit			&Implicit		&Implicit\\
				&internal		&internal			&external			&volatile		&external\\
			&{\em (references)}	&{\em (references)}	&{\em (connectors)}	&{\em (invocations)}	&{\em (roles)}\\
\hline\hline
{\sl Commu-}		&				&				&				&			&\\
{\sl nication}		&				&				&				&			&\\
\hline
{\sl Receiver(s)}	&Point to point		&Point to point		&Multi-point		&Multi-point	&Multipoint\\
{\sl designation}	&explicit			&explicit			&explicit			&dynamic		&explicit\\
				&				&			&or implicit		&{\em (discovery}	&or implicit\\
				&				&			&{\em (publish-}	&{\em and selection)}	&{\em (role}\\
				&				&				&{\em subscribe)}	&			&{\em designation)}\\
\hline
{\sl Data}			&Bi-			  	&Uni-			&Bi- or uni- 		&Bi- or uni-	&Uni-\\
{\sl Transfer}		&directional		&directional		&directional		&directional	&directional\\
				&{\em (value return)}		&			&{\em (events)}		&			&direct or indirect\\
				&				&				&				&			&{\em (via environment)}\\
\hline
{\sl Synchro-}		&Synchronous		&Asynchronous		&Synchronous		&Synchronous	&Asynchronous\\
{\sl nization}      	&		        		&	                         &or                  		&or                  	&or\\
				&				&				&asynchronous		&asynchronous	&protocol\\
\hline
\end{tabular}
\end{footnotesize}
\caption{Nature of the coupling}
\label{table:flexibilite:couplage}
\end{center}
\end{table}

%\begin{figure}
%\includegraphics[width=12cm]{evolution-couplage-en.png}
%\caption{Coupling flexibility evolution}
%\label{figure:evolution:flexibilite:couplage}
%\end{figure}

%STOP
%7 December 2021

\section{Abstraction Level}
\label{section:abstraction}

The history of programming begins with concepts very close to the machine (instructions, integers, etc.),
then progressively identifies
%a number of abstractions increasingly
some higher level {\em abstractions}
(procedure, function, data structure,
semaphore, process,
object, message, component, model, etc.).
The concepts of agent and organization continue this evolution towards more abstraction
as well as towards more explicit knowledge.
%but also towards more explicit.
%Let us consider the evolution of manipulated data and in particular exchanged.

\subsection{From Data to Concepts}

The transition from primitive data types to abstract data types
allows the modeling and naming of arbitrary classes of objects.
%, representing actual objects of the planned application.
Object-based programming introduces some major evolution step,
with objects
modeling and representing (reifying) conceptual or physical objects
of the application domain considered
\cite{perrotobjets98}.
%(see, e.g., a review of their successes,
%failures and prospects
%is proposed
%in
%\cite{introobjetsjfp04}).
%\cite{hommage:jfp:04}).
In other words, we moved from data to {\em concepts}.
Agents will extend this evolution with an explicitation of the domain (including human) {\em knowledge}.
Cognitive agents introduce the notion of {\em mental state},
inherited from symbolic artificial intelligence (see, e.g., \cite{aopshoham}),
with some symbolic representation of cognitive concepts,
such as: belief, goal, desire, intention, etc.
Furthermore, such internal knowledge of an agent can be communicated
to other (external) agents,
e.g., communication of beliefs, plans or/and intentions,
in order for agents to learn about each others or/and coordinate their actions.
Agents can also reason about the {\em context}
(as coined in
\cite{coutaz:context:05},
``context is key'')
%\footnote{\cite{chen:kotz:context:report:2000}
%	identifies four basic types of {\em context}:
%	{\em computational} context (i.e. state of resources of the device and of the network),
%	{\em user} context (i.e. persons, places, or/and objects),
%	{\em physical} context (e.g., luminosity, noise, or/and temperature)
%	and {\em temporal} context (e.g., hour, day, or/and period of the year).}
for {\em context-aware} applications such as ambient intelligence.
%\cite{ambient:intelligence:managing:context:book:2009}.
(\cite{chen:kotz:context:report:2000}
identifies four basic types of context:
{\em computational} context (i.e. state of resources of the device and of the network),
{\em user} context (i.e. persons, places, or/and objects),
{\em physical} context (e.g., luminosity, noise, or/and temperature)
and {\em temporal} context (e.g., hour, day, or/and period of the year).)

The object-oriented discipline of message sending also provides some self-documentation,
as the subject and the request type are specified explicitly.
Agent communication languages raise further the explicitness of information and knowledge.
Indeed, information that had remained implicit (and hidden) in object-oriented
and component-based applications
-- such as intention of communication, coordination logic, plans, etc. --
and remained in the mind of the programmer,
become explicit and thus better document the program.
Moreover, this information could also be used by the agents themselves
(for example to coordinate, reason about communication failures, replan, reorganize, etc.).

\subsection{Reification}

An additional approach, transversal to the type of abstractions proposed (objects, components\ldots, agents),
%and complementary to the process of abstraction,
is {\em reification}.
It is the process by which an abstract concept about a computer program
is turned into an explicit entity created in the programming language.
In other words, something that was previously implicit and unexpressed is explicitly formulated at the level of the language
(thus often coined as ``making something a first-class citizen''),
and therefore made available to {\em inspection} and {\em manipulation}.
The Lisp programming language has been a true pioneer, with its uniform vision of considering programs as data.
This has been developed further in the Smalltalk programming language,
which reifies various types of program and implementation entities,
such as messages, contexts, classes,
as actual Smalltalk objects.
Static concepts, e.g., classes, are reified as permanent objects.
(A class is thus an instance of a class, usually named a {\em metaclass}.)
Computational concepts, like stack contexts and messages, are only reified on demand (e.g., in case of errors), for obvious efficiency reasons.
The inverse operation, making a reified information (back) into an actual implementation, is named {\em reversion}, or {\em reflection}.
An example in Smalltalk occurs in case of an error:
the interactive debugger opens up and allows inspection of a reified context of the currently stopped computation.
Once correcting the error, the debugger reinstalls the corrected computational context and resumes computation.

In addition, in some languages, an entity already explicit at the programming level,
e.g., an object, may gain some explicit representation of some of its implementation characteristics,
usually coined as its (or one of its) {\em meta-object}.
Such types of self-described and introspective languages or architectures are usually named either {\em reflective architectures} or {\em meta-level architectures}.
Indeed, this way of opening up implementations (into manipulable abstractions) in order to make them adaptable at some high level of abstraction
turned out being very useful.
See, e.g.,
\cite{reflection:2:99} for some survey of reflective, meta-level and/or meta-object architectures,
\cite{class:ijcai:87} and \cite{cointe:objvlisp:oopsla:87}
for an example of minimal reflective object-oriented architecture,
and \cite{art:mop:91} for a very developed one.

\subsection{Interoperability Languages}
\label{section:abstraction:interoperability:languages}

Let us now examine {\em interoperability middleware},
which specifies and standardizes the exchange of information.
CORBA object-oriented middleware designed by OMG \cite{corba}
standardizes,
through an {\em interface description language} (IDL),
the types of data exchanged.
The analogue for agents further refines the way information is exchanged.
The IDL of CORBA is substituted
%(actually not exactly,
%more subtle that,
%as
%it will be
(see details
%explained
in next paragraph)
by a more general agent communication language
(ACL).
In addition to the specific content of the message, an ACL communication can specify:

\begin{itemize}

\item {\em performative}:
some symbolic designation of the {\em intention} of the communication
(e.g., inform, deny, recruit, etc.);

\item {\em content description language}:
the language used to describe the content.
It can be some programming language (e.g., Java) or some knowledge representation language
(e.g., KIF, or SL \cite{fipaacl});

\item {\em ontology}:
the ontology(s)
(i.e. some representation of a set of concepts, their properties and their relations)
of the concepts referred to by the message
(e.g., some standard ontology about transport and tourist services,
for some electronic travel agency application);

\item {\em protocol}:
the protocol used for the communication
(e.g., a call for proposals, named FIPA-Contract-Net, see Figure~\ref{figure:cnp}).

\end{itemize}

It should be noted
that CORBA and ACL do not actually play exactly the same roles \cite{automatedcomponentconfiguration}.
CORBA, through its IDL, provides some standard for specifying the interfaces (signatures) of objects and components.
It also provides mappings (named {\em projections}) of this IDL in different programming languages
(e.g., Java, Smalltalk, C++, etc.).
Therefore, CORBA can automatically generate {\em implementation skeletons} for the calling party code
and for the called party code,
and  thus ensure the translation and transfer of data.
An ACL does not offer some standard for specifying interfaces of agents,
but offers {\em instead} some general standard for specifying various properties of communication between agents,
which is different.
As listed above, ACL standardizes various properties such as intention, ontology and protocols.
The first historically is KQML \cite{kqml},
followed by FIPA ACL \cite{fipaacl}.

\subsection{Organizational Design}

It is also important to highlight the preponderant role of the {\em design} of multi-agent systems.
It is guided by the {\em organization of work} (through concepts such as organization, role, dependence, and {\em norms})
and by {\em knowledge} (mind states, such as belief and intentions),
rather than by the {\em operational means} for achieving this work,
which corresponds to the traditional procedural approach of programming (through data and procedures).
Multi-agent {\em methodologies} (e.g., such as the pioneering Cassiop\'ee \cite{cassiopee})
often start with some analysis of organizations, roles and their dependencies,
while considering separately (and later) implementation questions
(such as: which agents will fulfill the roles, depending on what decomposition of tasks).
Some agent-oriented design can then be carried out (implemented) in some multi-agent architecture,
or through objects, actors, or/and components,
the agent level not always appearing completely at the implementation level.
However,
keeping abstractions,
such as agents and organizations,
as entities explicitly represented at the {\em execution level},
offers of course possibilities of dynamic manipulation by the programmer,
but above all by the entities (agents and organizations) themselves, thus offering possibilities of self-adaptation and self-organization
(see, e.g., the organizational model MOISE \cite{moise+}).

Finally, in the evolution and the elevation of programming abstractions,
%as illustrated in Figure~\ref{figure:evolution:abstraction},
we also need to mention about {\em model driven engineering},
such as the {\em model driven architecture} (MDA) proposed by the OMG \cite{mda}),
as a modeling level for the partial automation of the construction of applications.
(For more details, see, e.g., the companion chapter by Jean-Marc J\'ez\'equel on modeling.)
Note that this line of research is somehow orthogonal to a specific programming model (object-oriented, component-based, agent-oriented, etc.).
There are efforts to couple multi-agent programming and model engineering,
see, e.g., \cite{mdasma}.

The vertical axis of Figure~\ref{figure:evolution:3} illustrates the evolution of abstraction level
within our frame of reference.

%\begin{figure}
%\includegraphics[width=12cm]{evolution-abstraction-en.png}
%\caption{Abstraction level evolution}
%\label{figure:evolution:abstraction}
%\end{figure}

%\begin{figure}
%\includegraphics[width=12cm]{evolution-3-en.png}
%\caption{Programming evolution}
%\label{figure:evolution:3}
%\end{figure}

\section{Conclusion}
\label{section:conclusion}

%We are witnessing a dual movement of two communities.
Due to the increasing needs for auto-adaptation
of future distributed applications (such as, e.g., Internet of Objects),
models of software components and software architectures are gradually gaining
in terms of abstraction as well as in (self) adaptation and reconfiguration capacities
(see, e.g., \cite{automatedcomponentconfiguration},
and also IBM proposal for autonomic systems
\cite{able}).
They get inspiration from multi-agent systems abstractions,
while often relying on light-weight infrastructures such as, or inspired by, web services.
The technology of web services is indeed simpler and lighter to implement and to deploy than some distributed component models (such as, e.g., CORBA),
as current web infrastructure is sufficient.
Web services provide the specification of the coordination between services (named choreography)
although it does not yet reach the level of sophistication of multi-agent systems
(on this topic, see, e.g., a comparative analysis of web services and agents \cite{paynewebservicesagents08}).
Additional abstractions from distributed programming
(such as, e.g., replication, groups and consensus, to manage fault-tolerance)
and/or ambient programming
(e.g., ambient references \cite{van:curtsem:ambienttalk:oopsla:06},
to
%be able to communicate with remote services in a context of
manage
volatile connexions),
are also needed to deal with distribution, fault-tolerance, volatility and uncertainty.
(Note that distribution was not the focus of this study, for more details, see, e.g., the companion chapter by Michel Raynal on distributed programming.)
%(and self-adaptation and self-reconfiguration).
%Simultaneously,
%%due to their success, and therefore industrial-level requirements,
%multi-agent models and platforms have progressed in terms of {\em methods} (methodologies) and development and deployment tools.
%%The frontier between the approaches thus gradually becomes more thin.
%However, cultural specificities sometimes lead to some ignorance about respective works.
%One of the objectives of this analysis is to humbly contribute to clarify and articulate various programming abstractions and their evolution
%%respective programming abstractions and issues from these communities,
%and thus to favor mutual awareness and possible cross-fertilization.

%Finally, we must recall the presence of a recent but important player in the game
%% that already exists between components and agents,
%namely
%these are the architectures based on services and a natural embodiment in the form of
%{\em web services}\footnote{Although
%	discussed	during our comparative analysis,
%	we have voluntarily restricted the description of concepts and works
%	in terms of service-oriented architectures, and in particular web services.
%	In addition, we also have that the current, however important, and relatively transverse,
%	of {\em model-driven engineering} (MDE),
%	as proposed by the OMG \cite{mda}).
%	This analysis actually initially focused on the relationships between software components and multi-agent systems,
%	resulting in an already extensive chapter.}.
An important stake is therefore to be able to integrate
and reuse, as much as possible,
respective abstractions and experience from various programming models and communities.
However, cultural specificities sometimes lead to some ignorance about respective works.
One of the objectives of this analysis was therefore
to (humbly) contribute
%to clarify various programming abstractions and their respective evolution and articulation,
to highlight some of the basic forces
motivating the progress of programming abstractions,
%%respective programming abstractions and issues from these communities,
in order
to favor mutual awareness, as well as possible cross-fertilization\footnote{In that respect,
	the last section of an article in french \cite{composants:agents:tsi:14} which was the basis for this chapter
	identifies various potential mutual cross contributions between software components and multi-agent systems:
	from agents to components, e.g., by using mapping and negotiation techniques to assist the assemblage of components;
	and from components to agent(s), e.g., to structure and modularize its architecture.},
and to provide some inspirational seeds about future programming abstractions.

%\acknowledgements{%
\subsection*{Acknowledgements}

The premises of this study go back to an interview that we conducted with Les Gasser
on the relationship between objects and agents,
%\cite{briotinterviewgasserieeeconcurrency98},
published in a special series on actors and agents \cite{briotinterviewgasserieeeconcurrency98}.
We thank him for his pioneering and fundamental contribution to this reflection
and we dedicate this article to his memory.
We also would like to thank Pierre Cointe and Jean-Fran\c{c}ois Perrot
for having formed us to the magic of programming.
%and for their enlightening contributions and for the chance of our collaboration.
%We also thank Jean-Fran\c{c}ois Perrot for his enlightening contributions on the concepts and practice of programming.
%(see for example \cite{perrotobjets98} and \cite{introobjetsjfp04}).
%We dedicate this article to them.

\bibliography{components-agents-briot}
\bibliographystyle{spmpsci}

\end{document}